\documentclass[acmsmall,screen=true]{acmart}

\usepackage{amsmath,amsfonts}
\usepackage{booktabs}
\usepackage{calligra}
\usepackage{color, colortbl}
\usepackage{courier}
\usepackage{csvsimple}
\usepackage{enumitem}
\usepackage{fancybox}
\usepackage{fontenc}
\usepackage{graphicx}
\usepackage{listings}
\usepackage{longtable}
\usepackage{lscape}
\usepackage{makecell}
\usepackage{marvosym}
\usepackage{moreverb}
\usepackage{multicol}
\usepackage{multirow}
\usepackage{pifont}
\usepackage{rotating}
\usepackage{setspace}
\usepackage{caption}
\usepackage{subfigure}
\usepackage[most]{tcolorbox}
\usepackage{threeparttable}
\usepackage{tikz}
\usepackage[normalem]{ulem}
\usepackage{url}
\usepackage{soul}
\usepackage{wasysym}
\usepackage{svg}
\usepackage{textcomp}
\usepackage{xcolor}
\usepackage{wrapfig}
\usepackage{pdflscape}
\usepackage{hyperref}

\usepackage{array}
\usepackage{balance}
\usepackage{microtype}

\newcommand{\intuition}[1]{
\begin{tcolorbox}[colback=white,boxrule=1pt,top=0pt,bottom=0pt,left=1pt,right=2pt,top=2pt,bottom=2pt]
\em #1
\end{tcolorbox}
}

\usepackage{makecell}
\usepackage{enumitem}
\usepackage{algorithm}
\usepackage[noend]{algpseudocode}
\usepackage{multirow}
\usepackage{caption}
\usepackage{xspace}

\newcommand{\Chain}{MulChain\xspace}

\begin{document}

\title{\Chain: Enabling Advanced Cross-Modal Queries in Hybrid-Storage Blockchains}

\author{Zhiyuan Peng}
\authornote{Equal contribution.}
\affiliation{%
  \institution{Shanghai Jiao Tong University;}
  \institution{The State Key Laboratory of Blockchain and Data Security, Zhejiang University;}
  \institution{Shanghai Jiao Tong University (Wuxi) Blockchain Advanced Research Center}
  \country{China}
  }
\email{pzy2000@sjtu.edu.cn}

\author{Xin Yin}
\authornotemark[1]
\affiliation{%
  \institution{The State Key Laboratory of Blockchain and Data Security, Zhejiang University}
  \city{Hangzhou}
  \country{China}
  }
\email{xyin@zju.edu.cn}

\author{Gang Wang}
\affiliation{%
  \institution{Northeastern University}
  \city{Shenyang}
  \country{China}
  }
\email{1910636@stu.neu.edu.cn}

\author{Chenhao Ying}
\authornote{Corresponding authors.}
\affiliation{%
  \institution{Shanghai Jiao Tong University}
  \city{Shanghai}
  \country{China}
  }
\email{yingchenhao@sjtu.edu.cn}

\author{Wei Chen}
\affiliation{%
  \institution{Shanghai Jiao Tong University}
  \city{Shanghai}
  \country{China}
  }
\email{chenwei8@sjtu.edu.cn}

\author{Xikun Jiang}
\affiliation{%
  \institution{University of Copenhagen}
  \city{Copenhagen}
  \country{Denmark}
  }
\email{xikun@di.ku.dk}

\author{Yibin Xu}
\affiliation{%
  \institution{University of Copenhagen}
  \city{Copenhagen}
  \country{Denmark}
  }
\email{yx@di.ku.dk}

\author{Yuan Luo}
\authornotemark[2]
\affiliation{%
  \institution{Shanghai Jiao Tong University}
  \city{Shanghai}
  \country{China}
  }
\email{yuanluo@sjtu.edu.cn}

\begin{abstract}

With its decentralization and immutability, blockchain has emerged as a trusted foundation for data management and querying. 
Because blockchain storage space is limited, large multimodal data files, such as videos, are often stored offline, leaving only lightweight metadata on the chain.
While this hybrid storage approach enhances storage efficiency, it introduces significant challenges for executing advanced queries on multimodal data. 
The metadata stored on-chain is often minimal and may not include all the attributes necessary for queries like time range or fuzzy queries.
In addition, existing blockchains do not provide native support for multimodal data querying. 
Achieving this capability would necessitate extensive modifications to the underlying blockchain framework, even reconstructing its core architecture.
Consequently, enabling blockchains with multimodal query capabilities remains a significant problem, which necessitates overcoming the following three key challenges: (1) Designing efficient indexing methods to adapt to varying workloads that involve frequent insertions and query operations; (2) Achieving seamless integration with existing blockchains without altering the underlying infrastructure; (3) Ensuring high query performance while minimizing gas consumption. 
To address these challenges, we propose \textit{\Chain}, a novel middleware architecture to enable smooth integration with existing blockchains. 
At the core of \Chain is the BHashTree, a flexible data structure that dynamically switches between tree and hash nodes based on workload characteristics, ensuring efficient insertion and query operations. 
Furthermore, the middleware provides standardized interfaces for blockchain systems, unifying query methods across different platforms. 
Finally, Experiments on Ethereum and FISCO-BCOS demonstrate that \Chain outperforms the state-of-the-art vChain+ by up to 78.12X in query performance and up to 99.49\% reduction in verification object size, respectively.

\end{abstract}



\maketitle

\section{Introduction}
Blockchain is regarded as a promising technology for data storage and management in Web 3.0 and the metaverse~\cite{gadekallu2022blockchainmetaversereview}. 
As the digital world evolves at an unprecedented pace, Web 3.0 and the metaverse are becoming integral parts of this new era of online interaction.
Blockchain technology is at the heart of these advancements, playing a crucial role in enabling decentralization, trust, and ownership in virtual environments.
It enables decentralized ownership through NFTs, allowing users to truly own and control their digital assets.    
Blockchain also supports portable and sovereign digital identities, enhancing privacy and security in virtual environments. 
Smart contracts automate transactions and governance, while cryptocurrencies facilitate digital economies. 
Moreover, blockchain's interoperability enables seamless asset and data transfer across different virtual worlds, paving the way for a more immersive and interconnected digital future.

Due to the limited storage resource of blockchains, a hybrid-storage strategy is typically employed for storing diverse content, where large volumes of multimodal data (e.g., videos, audio, and images) are stored off-chain, while only transactions and metadata are retained on-chain~\cite{arer2022efficient,liu2024mpv,MSTDB2023,zhu2019sebdb}. 
For example, in an NFT art marketplace, the metadata (e.g., IDs, titles, creators, ownership history) of NFTs is stored on-chain, while the actual multimedia contents (e.g., high-resolution images and promotional videos) are stored on off-chain storage (e.g., IPFS~\cite{benet2014ipfscontentaddressed}). 
By utilizing data from hybrid-storage blockchains, users can view and verify the contents associated with NFT artworks~\cite{grtandnfts23}.
In a DeFi application, on-chain data provides information about users' liquidity, loan repayments, etc., while off-chain data includes traditional credit scores and income records. By combining on-chain transaction history with off-chain credit data, a user's financial reliability can be evaluated.

As the demand for storing multimodal data in blockchain systems grows rapidly, there is a pressing need for storage solutions that can handle more sophisticated and comprehensive query operations~\cite{V2FS2024}. For example, querying for NFT details often requires combining on-chain metadata with off-chain multimedia data, involving complex operations like fuzzy and string matching. Similarly, calculating users' total liquidity in decentralized finance (DeFi) applications may require filtering both on-chain and off-chain credit scores through range queries.
Currently, blockchain systems typically support two strategies for verifiable query processing. 
The first strategy involves blockchain-based databases, which can handle only basic query types, such as boolean range queries~\cite{chang2023anole,wang2022vchain+,xu2019vchain}. 
While simple, this approach is limited in query complexity and tends to be proprietary, making it hard to be compatible with other blockchain systems~\cite{V2FS2024}. 

The second strategy involves outsourcing queries to external databases, which can handle more complex queries, such as time range queries~\cite{antonopoulos2021sql,peng2020falcondb,yang2020ledgerdb,yue2022glassdb,zhang2017vsql,zhang2015integridb}. 
However, this approach introduces significant risks, particularly the potential for information leakage, as it relies on external storage. 
Furthermore, many existing solutions only support a single type of query, lacking compatibility with other query methods.

As the variety of stored data types increases—encompassing text, images, videos, and more—the need for efficient cross-modal data querying becomes increasingly important. 
Existing blockchain storage systems do not address this need, especially in terms of supporting both hybrid-storage queries and efficient cross-modal operations. 
In this paper, we propose a novel middleware architecture for hybrid-storage queries, designed to offer high compatibility, support a broad range of query operations, and enable efficient cross-modal data querying.

\begin{table*}
\centering
\caption{Comparison with Existing Query Systems}
\label{tab:Comparison}
\resizebox{\textwidth}{!}{
\begin{tabular}{cccccc}
  \toprule
  \textbf{Category} &
  \textbf{Name} &
  \makecell{\textbf{Query} \\ \textbf{Language}} &
  \makecell{\textbf{Blockchain} \\ \textbf{Compatible}} &
  \makecell{\textbf{Multiple} \\ \textbf{Advanced} \textbf{Queries}} &
  \makecell{\textbf{Off-Chain}\\ \textbf{Compatible}} \\ 
  \midrule
  \multirow{6}{*}{\textbf{\begin{tabular}[c]{@{}c@{}}Outsourced \\ Database\end{tabular}}} &
  IntegriDB~\cite{zhang2015integridb} &
  Semi-SQL &
  N/A &
  \textcolor{red}{\textbf{\ding{55}}} &
  \textcolor{red}{\textbf{\ding{55}}} \\
 &
  FalconDB~\cite{peng2020falcondb} &
  Semi-SQL &
  N/A &
  \textcolor{red}{\textbf{\ding{55}}} &
  \textcolor{red}{\textbf{\ding{55}}} \\
 &
  vSQL~\cite{zhang2017vsql} &
  SQL &
  N/A &
  \textcolor{red}{\textbf{\ding{55}}} &
  \textcolor{red}{\textbf{\ding{55}}} \\
 &
  SQL Ledger~\cite{antonopoulos2021sql} &
  SQL &
  N/A &
  \textcolor{red}{\textbf{\ding{55}}} &
  \textcolor{red}{\textbf{\ding{55}}} \\
 &
  LedgerDB~\cite{yang2020ledgerdb} &
  Read &
  N/A &
  \textcolor{red}{\textbf{\ding{55}}} &
  \textcolor{red}{\textbf{\ding{55}}} \\
 &
  GlassDB~\cite{yue2022glassdb} &
  Read &
  N/A &
  \textcolor{red}{\textbf{\ding{55}}} &
  \textcolor{red}{\textbf{\ding{55}}} \\ 
  \midrule
  \multirow{6}{*}{\textbf{\begin{tabular}[c]{@{}c@{}}Blockchain \\ Database\end{tabular}}} &
  vChain~\cite{xu2019vchain}, vChain+~\cite{wang2022vchain+} &
  Boolean Range &
  \textcolor{red}{\textbf{\ding{55}}} &
  \textcolor{red}{\textbf{\ding{55}}} &
  \textcolor{red}{\textbf{\ding{55}}} \\
 &
  GEM$^2$-tree~\cite{GEM^2-Tree} &
  Range &
  \textcolor{red}{\textbf{\ding{55}}} &
  \textcolor{red}{\textbf{\ding{55}}} &
  \textcolor{red}{\textbf{\ding{55}}} \\
 &
  LVQ~\cite{dai2020lvq} &
  Membership &
  \textcolor{red}{\textbf{\ding{55}}} &
  \textcolor{red}{\textbf{\ding{55}}} &
  \textcolor{red}{\textbf{\ding{55}}} \\
 &
  The Graph~\cite{thegraph2022} &
  GraphQL &
  \textcolor{green}{\textbf{\ding{51}}} &
  \textcolor{red}{\textbf{\ding{55}}} &
  \textcolor{red}{\textbf{\ding{55}}} \\
 &
  V2FS~\cite{V2FS2024} &
  Various Types &
  \textcolor{green}{\textbf{\ding{51}}} &
  \textcolor{green}{\textbf{\ding{51}}} &
  \textcolor{red}{\textbf{\ding{55}}} \\
 &
 \textbf{\Chain(ours)} &
  SQL &
  \textcolor{green}{\textbf{\ding{51}}} &
  \textcolor{green}{\textbf{\ding{51}}} &
  \textcolor{green}{\textbf{\ding{51}}} \\ 
  \bottomrule
\end{tabular}%

}
\end{table*}

However, designing a system that supports hybrid storage and cross-modal data querying presents a range of new challenges, as highlighted in Table~\ref{tab:Comparison}~\cite{MSTDB2023}. 
First, ensuring compatibility with existing blockchain frameworks is of paramount importance, as popular networks like Bitcoin~\cite{nakamoto2008bitcoin} and Ethereum~\cite{wood2014ethereum} boast significant market presence. 
Modifying their core structures would be impractical and disruptive~\cite{V2FS2024}.
Second, the system must accommodate a wide array of SQL queries to meet diverse client requirements, demanding robust and adaptable query processing capabilities. 
Third, real-time query performance poses a major hurdle due to the inherent latency and throughput limitations of blockchain networks, necessitating the development of optimized mechanisms for timely data retrieval. 
Finally, supporting multimodal data queries is crucial, as Decentralized Applications (DApps) often handle a variety of data types and formats, requiring an integrated approach for seamless querying.

To address these challenges, we propose \textit{\Chain}, a novel middleware architecture to enable blockchain compatibility and advanced queries for multimodal data. 
\Chain is a modular and pluggable middleware that orchestrates seamless integration between existing blockchain systems (e.g., Ethereum) and off-chain storage solutions (e.g., IPFS). 
Based on this, we present a gas-efficient BHashTree alongside a high-performance trie designed to maximize efficiency and enhance query capabilities.
Note that both index structures are verifiable. 
This approach differs from previous indexing methods and supports various advanced query functions. 
Furthermore, we provide a handful of advanced SQL query primitives supporting time range and fuzzy queries. To make a fair comparison with \Chain, we reimplemented vChain+ to support cross-modal queries on hybrid-storage blockchains. 
Experimental results show that \Chain surpasses state-of-the-art accumulator-based vChain+ by up to 22.67X speedup in simple queries, up to 78.13X speedup in time range queries, 30X speedup in fuzzy queries and reduces the VO size by up to 99.49\%.
In summary, the key contributions of this paper include:

\begin{itemize}[leftmargin=*]
    \item We propose \Chain, a pluggable middleware architecture to enable advanced cross-modal queries without modifying existing blockchains. 
    \Chain provides standardized interfaces and unifies query methods across different platforms, ensuring interoperability between blockchains and off-chain storage solutions. To the best of our knowledge, this is the first work to enable advanced cross-modal queries in hybrid-storage blockchains.

    \item We design and implement BHashTree, a flexible index structure that dynamically adapts between tree and hash nodes based on workload characteristics. Also we implement trie to support high performance fuzzy query. These data structures empower the system to support multiple cross-modal query types.

    \item We experiment on two popular blockchains (Ethereum and FISCO BCOS) and one off-chain storage (IPFS). 
    Experimental results show that \Chain achieves up to 78.12X speedup in on-chain queries and reduces the VO size by 99.49\% compared to the SOTA vChain+.

    \item To support the open science community, we publish the studied dataset and source code of our model with supporting scripts on GitHub ({\url{https://github.com/pzy2000/MulChain}}), which provides a ready-to-use implementation of our model for future research about comparison.
\end{itemize}

\section{Motivation}

\subsection{A Motivation Example}

\begin{figure}[htbp]
    \centering
    \includegraphics[width=\linewidth]{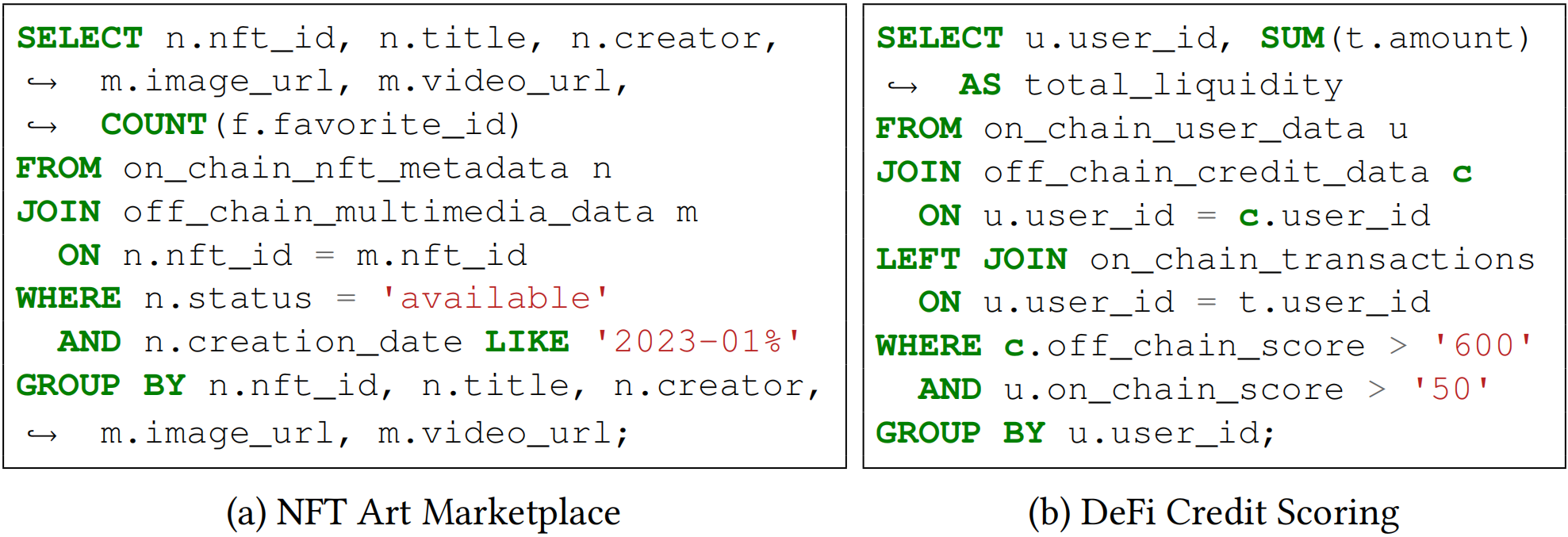}
    \caption{Querying Multimodal Data on Hybrid-Storage Blockchain}
    \label{fig:motivation}
\end{figure}

As the requirement of storage for multimodal data is increasing exponentially in blockchain, the storage system is required to offer more sophisticated and comprehensive data queries~\cite{V2FS2024}. In Fig.~\ref{fig:motivation}, we present two representative queries that highlight the application of multimodal data in hybrid-storage blockchains. Fig.~\ref{fig:motivation}(a) illustrates a query designed for an NFT art marketplace, where it retrieves metadata from on-chain sources alongside multimedia data, such as images and videos, stored off-chain. This query aggregates user interactions by counting the number of favourites per NFT, ensuring the retrieval of relevant, available NFTs from January 2023, and combining both on-chain and off-chain information.

Fig.~\ref{fig:motivation}(b) showcases a query used in decentralized finance (DeFi) credit scoring, which integrates on-chain user data with off-chain credit information. By joining these datasets, the query computes the total liquidity of users with a credit score above a certain threshold, utilizing both transaction data and off-chain credit scores to derive a comprehensive assessment of the users' financial behaviour and standing. The use of both on-chain and off-chain data enables a more robust, holistic view of user activity and financial trustworthiness.


\begin{figure}[htbp]
    \centering
    \includegraphics[width=\linewidth]{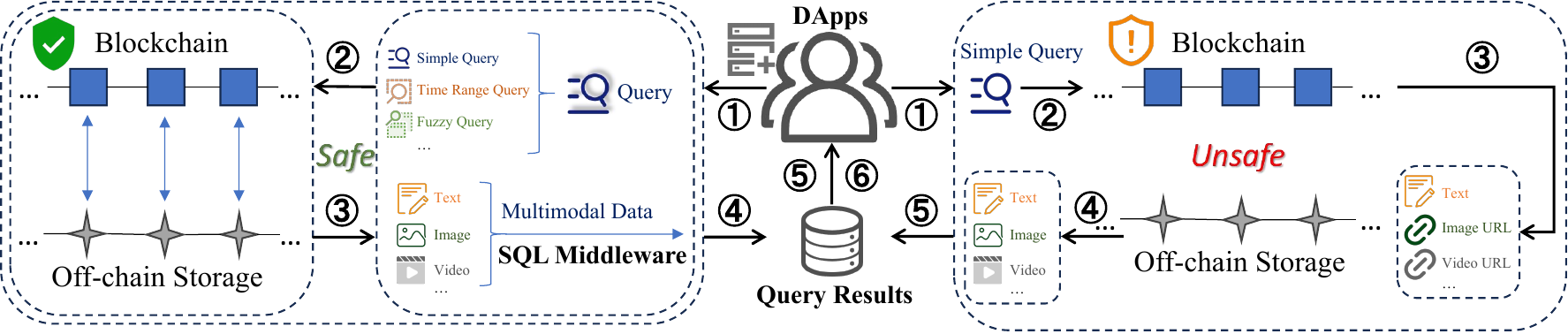}
    \caption{Traditional Hybrid Storage and \Chain Middleware}
    \label{fig:motivation_1}
\end{figure}

\textbf{Observation 1.} 
As shown on the right side of Fig.~\ref{fig:motivation_1}, querying URLs for data such as images and videos through simple queries and accessing them directly poses significant security risks.
This is primarily because it bypasses necessary security and privacy protocols inherent in hybrid-storage blockchain systems.
Specifically, fetching multimodal data through URLs could expose sensitive data to unauthorized access, compromising the integrity and confidentiality of both on-chain and off-chain resources. In such cases, malicious actors could exploit these URLs to directly manipulate or retrieve information from off-chain storage systems, undermining the trust in the blockchain's decentralized security framework.

To mitigate this risk, the process of querying multimodal data must be controlled and mediated through a secure middleware layer (as shown on the left side of Fig.~\ref{fig:motivation_1}).
In this approach, the query first reaches the middleware, which acts as a trusted intermediary between the user and the blockchain. 
The middleware employs a SQL query parser to analyze and parse the incoming request, ensuring that only valid and authorized queries are processed. Upon successful parsing, the middleware then interacts with the blockchain through a smart contract, executing the query on the blockchain and retrieving the relevant results. 
These results typically include both on-chain metadata, such as text data, and off-chain resources, such as multimedia content identified by their unique content identifiers (CIDs).

Once the query results are obtained, the middleware further resolves the CIDs, using them to fetch the raw multimedia data (e.g., images and videos) from an InterPlanetary File System (IPFS)~\cite{benet2014ipfscontentaddressed} or other decentralized storage solutions. 
This ensures that the process of accessing and retrieving multimodal data remains secure, as the security of the middleware is guaranteed by the verifiable data structures in Section~\ref{sect:Structures}.

\textbf{Observation 2.} 
Existing blockchains offer two strategies for verifiable query processing.
\ding{182} The first approach involves blockchain-based databases, which typically support only basic queries, such as boolean range queries~\cite{chang2023anole,wang2022vchain+,xu2019vchain}. These systems are often self-designed, and while they can execute straightforward queries on on-chain data, they are not well-suited for handling complex, multimodal queries that involve both on-chain and off-chain data sources. 
Moreover, these blockchain-based databases are inherently closed systems, making it difficult to integrate them with other blockchain platforms or external storage solutions. As seen in Figure \ref{fig:motivation}(a) and (b), which feature queries that span both on-chain and off-chain data, such as NFTs with multimedia content (images and videos), such systems cannot handle the complexity or the cross-modal nature of these queries. This limitation is especially critical given that current blockchain storage methods fail to account for the necessity of supporting cross-modal data retrieval, including the secure fetching of multimedia files (e.g., images and videos) without exposing sensitive information through URLs.
\ding{183} The second strategy is outsource-based databases that enable more complicated queries~\cite{antonopoulos2021sql,peng2020falcondb,yang2020ledgerdb,yue2022glassdb,zhang2017vsql,zhang2015integridb}, such as time range queries. However, it introduces risks of information leakage due to reliance on external storage, and most of the existing methods support only a single query type and lack compatibility with others.
As illustrated in Figure \ref{fig:motivation}(a) and (b), the querying process involves data from multiple sources—on-chain metadata, off-chain multimedia data, and sometimes cross-modal relationships between them. For example, in the NFT art marketplace scenario, the query aggregates metadata from both the blockchain and off-chain multimedia storage to count user interactions with NFTs. In the DeFi credit scoring case, user data is drawn from both on-chain and off-chain sources, combining transaction data with credit scores. These queries not only require the ability to handle multimodal data but also need to perform cross-modal operations that seamlessly combine text, images, and video data in a secure, efficient manner. Unfortunately, existing blockchain storage strategies do not accommodate this growing need for cross-modal query support, particularly when security and privacy are critical considerations.
Due to these limitations, users are restricted in their ability to conduct comprehensive and diverse data queries, which hinders the efficient management and utilization of data in Web 3.0 and the metaverse.

\intuition{{\bf Intuition.}
Existing strategies respectively face the following limitations: (1) Lack of support for complex queries, and are difficult to be compatible with other blockchain systems; (2) Lack of support for cross-modal queries.  
}

\subsection{Key Ideas}
Based on the above observations, we aim to design a novel middleware architecture for hybrid-storage queries that offers high compatibility, supports diverse query operations, and enables efficient cross-modal data querying.

However, designing a system that supports hybrid-storage and cross-modal data querying introduces several new challenges~\cite{MSTDB2023}.
Firstly, maintaining compatibility with existing blockchain systems is crucial because popular blockchain networks (e.g., Bitcoin~\cite{nakamoto2008bitcoin} and Ethereum~\cite{wood2014ethereum}) have enormous market values, and any modifications to their underlying systems would be impractical and inconvenient~\cite{V2FS2024}.

Secondly, the system is required to support different types of SQL queries to meet clients' demands, necessitating flexible and comprehensive query processing solutions. Thirdly, real-time querying presents a challenge due to inherent latency and performance limitations in blockchain networks, requiring efficient mechanisms to enable timely data retrieval. 
Lastly, supporting queries over multimodal data is essential, as Decentralized Applications (DApps) often need to handle various data types and formats.

To address these challenges, we propose \textit{\Chain}, a novel middleware architecture to enable advanced queries for multimodal data.

\textbf{(1) Novel Middleware Architecture.}  
Traditional hybrid-storage blockchains store multimodal data in a two-step process: metadata (e.g., URL, hash value) is stored on-chain, and multimedia content (e.g., images and videos) is stored off-chain (e.g., IPFS). However, this architecture lacks compatibility with different blockchain systems and off-chain storage solutions, and it does not support complex, cross-modal queries. To address these issues, we propose \Chain, a novel middleware architecture that ensures compatibility across various blockchain systems (e.g., FISCO BCOS, Ethereum) and off-chain storage, while supporting advanced queries involving both on-chain and off-chain data.
Additionally, \Chain's secure middleware layer acts as an intermediary between the user and the blockchain, verifying queries and securely fetching off-chain data to prevent unauthorized access, ensuring the integrity of the entire system.

\textbf{(2) Verifiable BHashTree and Trie.}  
Efficient blockchain queries require handling multimodal data from diverse sources while ensuring security. To enable this, \Chain incorporates Verifiable BHashTree and Trie data structures. The BHashTree optimizes range and time-range queries by dynamically adapting between B+Tree and hash table representations based on query workload, ensuring efficient storage and retrieval in hybrid-storage scenarios. The Trie, on the other hand, excels at handling fuzzy queries where data may have variable formats or slight deviations, such as timestamps or addresses. This enables efficient retrieval even when exact matches are unavailable.

Both the BHashTree and Trie are integrated into \Chain to provide a verifiable, gas-efficient query mechanism for multimodal data, preserving blockchain security while supporting cross-modal queries. This combination allows for seamless retrieval and integration of on-chain metadata and off-chain multimedia data, enabling more powerful and flexible query processing. 

\section{System Architecture}

\subsection{Components}
\begin{figure}[htbp]
    \centering
    \includegraphics[width=.95\linewidth]{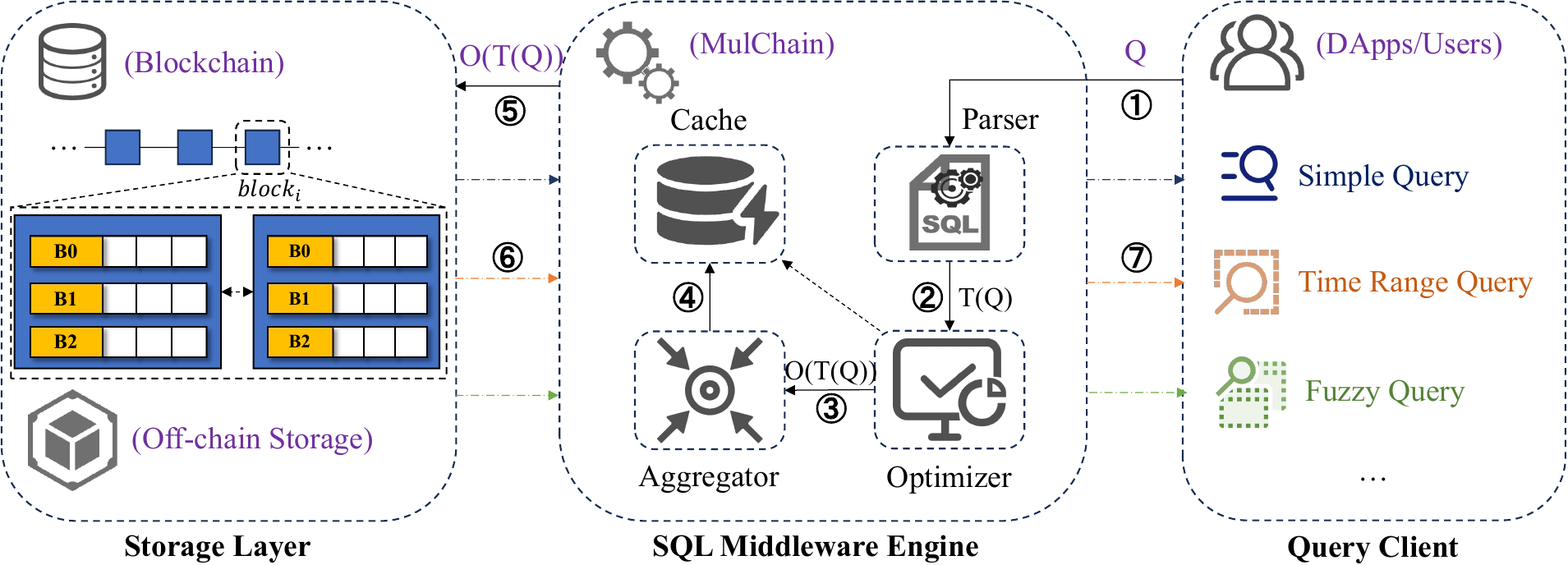}
    \caption{System Model of \Chain}
    \label{fig:overview}
\end{figure}

In \Chain, the integration of on/off-chain resources enables efficient cross-modal SQL queries. 
This design allows users to execute complex queries. 
By leveraging the SQL middleware engine, \Chain balances performance and scalability in blockchain environments.
\Chain has three parties (illustrated in Fig.~\ref{fig:overview}): \ding{182} \textbf{Storage Layer}, \ding{183} \textbf{SQL Middleware Engine}, and \ding{184} \textbf{Query Client}.

\subsubsection{Storage Layer}
The Storage Layer manages various multimodal data (e.g., text, images, and videos) by utilizing decentralized storage solutions (e.g., IPFS) to store large multimedia files off-chain while keeping essential metadata on-chain to guarantee data integrity and provenance. 
In the blockchain network~\cite{nakamoto2008bitcoin}~\cite{wood2014ethereum}, there are three types of nodes: full nodes, miner nodes, and light nodes. 
Full nodes maintain a complete copy of the blockchain ledger, ensuring data availability and validating transactions. 
Minor nodes contribute computational power to process transactions and add new blocks to the blockchain through consensus mechanisms, securing the network and maintaining its integrity. Light nodes store only essential information and rely on full nodes for data verification~\cite{V2FS2024}.

\subsubsection{SQL Middleware Engine}
The SQL Middleware Engine acts as a bridge between the query client and storage layer, the SQL middleware engine interprets high-level SQL queries into executable commands. Let a query \( Q \) be represented by a set of logical operations, \( Q = \{ L_1, L_2, \ldots, L_n \} \), where each \( L_i \) represents a logical operation that needs to be mapped to corresponding blockchain functions. The middleware engine uses a mapping function, \( T: Q \to C \), where \( C \) is the set of commands comprehensible by the blockchain network. Mathematically:

\begin{equation}
T(Q) = \{ T(L_1), T(L_2), \ldots, T(L_n) \}
\end{equation}

\subsubsection{Query Client}
The Query Client serves as the user interface for submitting SQL queries within the \Chain architecture. 
By interfacing directly with the SQL Middleware Engine, the Query Client delegates the interpretation and execution of high-level SQL queries to the middleware. 
This separation of concerns allows the Query Client to provide seamless query capabilities without storing or processing the entire distributed ledger. 
Consequently, users can perform complex cross-modal queries with reduced resource consumption, leveraging the scalability and performance benefits of the SQL Middleware Engine.

\subsection{Execution Flow}
\begin{figure}[htbp]
    \centering
    \includegraphics[width=.9\linewidth]{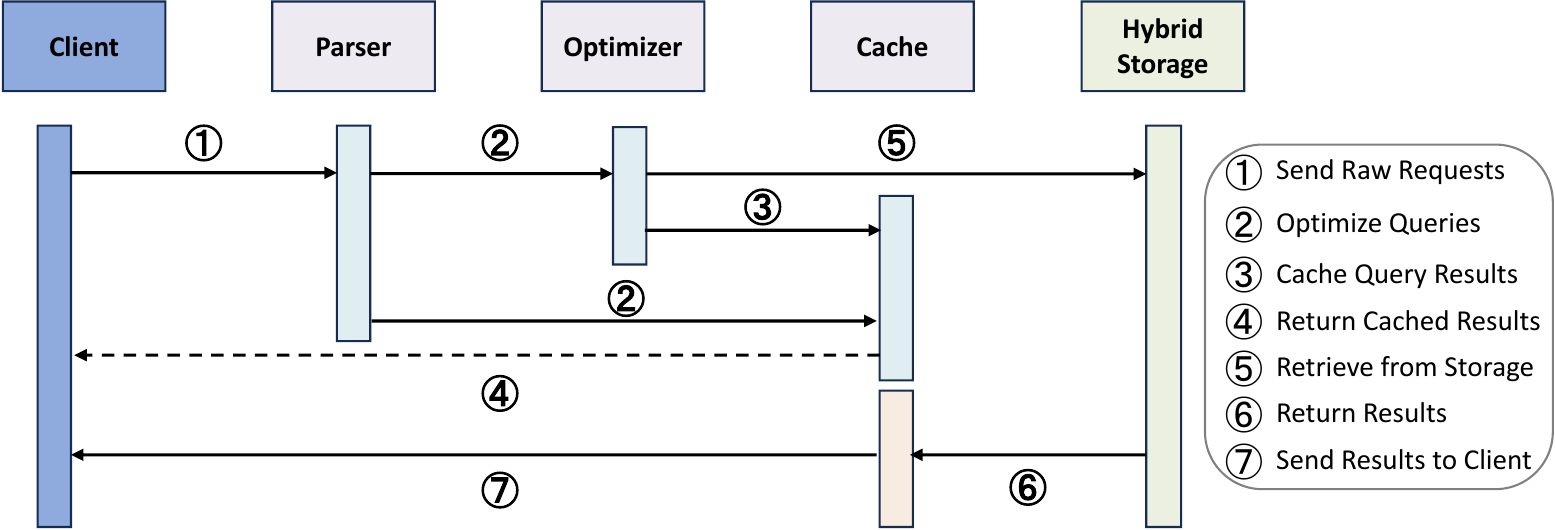}
    \caption{Execution Flow of \Chain}
    \label{fig:execution_flow}
\end{figure}

As shown in Fig.~\ref{fig:execution_flow}, the execution flow of \Chain is divided into the following steps:

\begin{enumerate}[leftmargin=*]
    \item Users initiate the process by submitting raw SQL queries \( Q \) through the User Interface Layer, which are then sent to the Parser for processing (\(\raisebox{-0.5ex}{\fontsize{14pt}{12pt}\selectfont\ding{192}}\)).

    \item The Optimizer refines \( Q \) into an execution plan \( P = \{p_1, p_2, \dots, p_n\} \) using dynamic programming, aiming to minimize the total computational cost \( C = \sum_{i=1}^n c(p_i) \), where \( c(p_i) \) represents the cost of each step (\(\raisebox{-0.5ex}{\fontsize{14pt}{12pt}\selectfont\ding{193}}\)).

    \item The Cache checks for previously computed results \( R_c \) matching \( Q \). 
    If \( R_c \neq \emptyset \), the cached results are returned to avoid redundant computation (\(\raisebox{-0.5ex}{\fontsize{14pt}{12pt}\selectfont\ding{194}}\)).
    If a match is found, the cached results \( R_c \) are directly delivered to the client (\(\raisebox{-0.5ex}{\fontsize{14pt}{12pt}\selectfont\ding{195}}\)).

    \item For uncached queries \( Q_u \), the Query Processing Unit retrieves the required data \( D = \{d_1, d_2, \dots, d_k\} \) by coordinating with both the blockchain and decentralized off-chain storage, ensuring consistency and data availability (\(\raisebox{-0.5ex}{\fontsize{14pt}{12pt}\selectfont\ding{196}}\)). 
    \item
    The retrieved data is integrated and formatted based on \( P \), generating the final results \( R \) for delivery (\(\raisebox{-0.5ex}{\fontsize{14pt}{12pt}\selectfont\ding{197}}\)).
    
    \item 
    Finally, the results \( R \) are returned to the client through the User Interface Layer (\(\raisebox{-0.5ex}{\fontsize{14pt}{12pt}\selectfont\ding{198}}\)).
\end{enumerate}


\section{Indexing Structures for Advanced Cross-Modal Queries}
\label{sect:Structures}

In this section, we present two novel indexing structures designed to enhance the efficiency of advanced cross-modal queries.
Specifically, a verifiable and gas-efficient BHashTree is utilized for time range queries, whereas a verifiable high-performance trie is employed for fuzzy queries. 
Furthermore, we delve into the analysis of how the shift from a traditional B\(+\)Tree to a BHashTree reduces time complexity.

\subsection{Verifiable Gas-efficient BHashTree}
\label{sect:BHashTree}

In high-intensity insertion scenarios, traditional tree-based indexes face severe scalability bottlenecks in time-series workloads, as the insertion of monotonically increasing timestamp keys causes high contention within small memory. 
In contrast, Hash tables achieve O(1) time complexity for queries and inserts, making them highly efficient for exact queries and insertions. 
However, they lack key order and cannot perform range queries or sorting efficiently~\cite{cha2023blink}.

\begin{figure}[htbp]
    \centering
    \includegraphics[width=.9\linewidth]{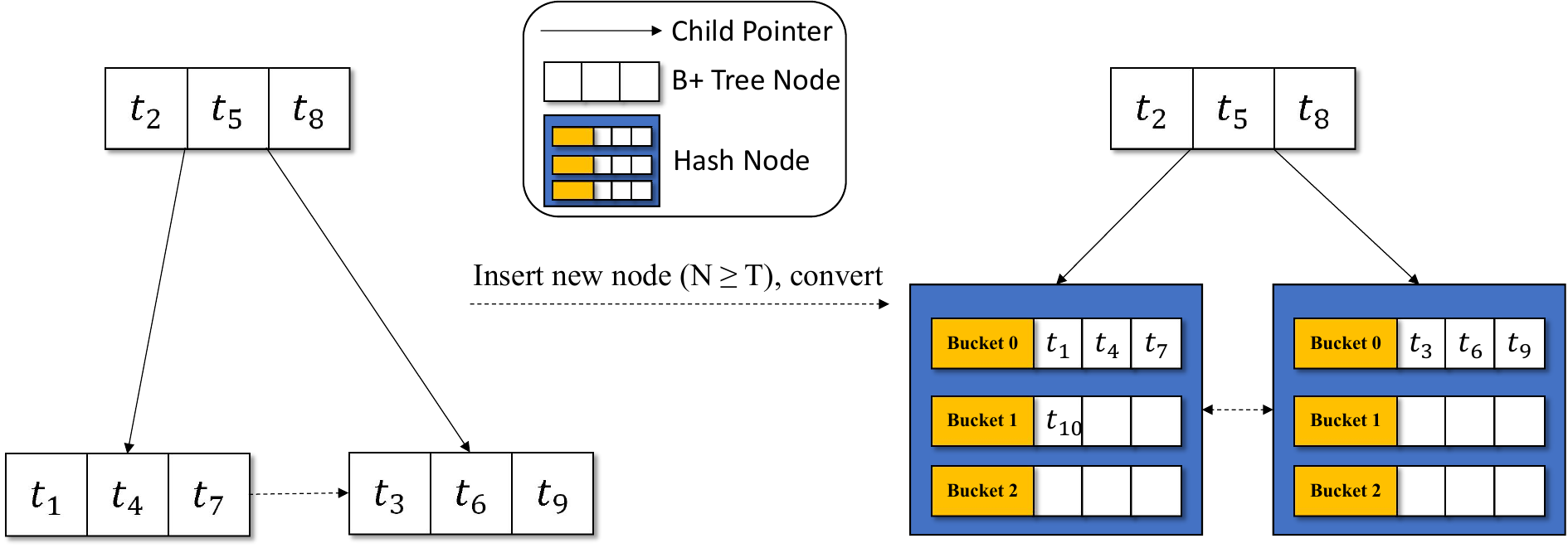}
    \caption{Dynamic Conversion of BHashTree. When the number of nodes reaches the predefined threshold $T$, \Chain converts B\(+\)Tree nodes to hash nodes.}
    \label{fig:Conversion}
\end{figure}

\begin{algorithm}[htbp]
\caption{Insertion Phase: Insert Data into BHashTree}
\label{Insert_Phase_BHashTree}
\begin{algorithmic}[1]
\Function{InsertToBHashTree}{entryId, timestamp}
    \State timeKey $\gets$ $\mathcal{C}$(\text{timestamp})
    \State rootNode $\gets$ BHashTreeNodes.rootId
    \If{rootNode.isLeaf \textbf{and} |rootNode.dEIds| $\geq$ T \textbf{and} \\
        \quad $\neg$ rootNode.isHashNode}
        \State rootNode.isHashNode $\gets$ \textbf{true}
        \State rootNode.fingerprint $\gets$ $\mathcal{C}$(\text{rootNode.dEIds}) \Comment{Compress fingerprints}
        \ForAll{entryId $\in$ rootNode.dEIds}
            \State key $\gets$ $\mathcal{F}$(rootNode.keys, entryId)
            \State rootNode.hashEty.key $\gets$ rootNode.hashEty.key $\cup$ \{entryId\}
        \EndFor
        \State rootNode.keys $\gets$ $\emptyset$
        \State rootNode.dEIds $\gets$ $\emptyset$
    \EndIf
    \If{rootNode.isHashNode}
        \State rootNode.hashEty.timeKey $\gets$ rootNode.hashEty.timeKey $\cup$ \{entryId\}
    \Else
        \State rootNode.keys $\gets$ rootNode.keys $\cup$ \{timeKey\}
        \State rootNode.dEIds $\gets$ rootNode.dEIds $\cup$ \{entryId\}
    \EndIf
    \State \Return True
\EndFunction
\end{algorithmic}
\end{algorithm}

To address these limitations, we propose BHashTree: a verifiable, gas-efficient index structure optimized for time-range queries. 
We modify the hash leaf node structure by employing a compressed fingerprint approach and fixed-size key-value pairs, thereby minimizing storage costs and ensuring efficient on-chain operations. 
We also adopt a dynamic adaptation strategy in which the system monitors insertion and scanning patterns through smart contract logic. This approach allows for the dynamic conversion of B\(+\)Tree nodes into hash nodes as scanning operations increase, optimizing storage and retrieval costs. 
As depicted in Fig.~\ref{fig:Conversion}, when the total number of B\(+\)Tree nodes reaches a predefined threshold \(T\) (set to 10 in our implementation), we iterate through the entire tree, transferring the values from tree nodes into hash nodes. 
Subsequently, we clear the keys and entryIDs of tree nodes to save storage.

To make BHashTree verifiable, we incorporate signed digests into the hash nodes. 
Following~\cite{pang2004authenticating}, we use a one-way hash function to compute the digest. 
Each leaf node stores a digest that encapsulates the key-value pair. 
Intermediate nodes maintain digests that propagate integrity up the tree. 
BHashTree generates a VO for time-range queries.
The VO includes digests for branches outside the queried range. 
This enables users to confirm result accuracy without accessing the entire structure. 
To preserve verifiability, we transfer digests for converted nodes without recalculating the whole tree. 
The root digest serves as a cryptographic anchor.
This allows efficient on-chain verification.
Algorithm \ref{Insert_Phase_BHashTree} and \ref{Query_Phase_Retrieve_Data} present the insertion and query phases of BHashTree.

\subsubsection{Insertion phase}
As shown in Algorithm \ref{Insert_Phase_BHashTree}, lines 4 to 12 depict the conversion when the number of B\(+\)Tree nodes reaches the threshold $T$. 
Lines 7 to 10 set the node to a hash node and reorganize existing entries into the hash mapping. 
Lines 13 to 17 manage the insertion of new entries into the appropriate structure based on the status of rootNode.

\subsubsection{Query phase}
As shown in Algorithm \ref{Query_Phase_Retrieve_Data},
the function initializes search parameters and prepares the result collection (lines 2 to 4). 
The recursive function \texttt{FindEntriesByBHT} (lines 8 to 25) traverses the BHashTree to find entries within the specified time range. 
For leaf nodes, if the node is a hash node (lines 11 to 15), it iterates over the keys within the range and collects the corresponding entries; 
if it is a B\(+\)Tree node (lines 16 to 20), it scans through the keys and selects entries that match the time range.
Lines 22 to 24 handle internal nodes by recursively searching child nodes whose key ranges overlap with the query range.

\begin{algorithm}[htbp]
\caption{Query Phase: Retrieve Data by BHashTree}
\label{Query_Phase_Retrieve_Data}
\begin{algorithmic}[1]
\Function{GetDataByTime\_BHash}{startTime, endTime}
    \State \textit{sK} $\gets$ $\mathcal{C}$(startTime), \textit{eK} $\gets$ $\mathcal{C}$(endTime), \textit{rEIds} $\gets$ $\emptyset$
    \State \textit{count, rEIds} $\gets$ \Call{FindEntriesByBHT}{rootId, sK, eK, rEIds, count}
    \State $D \gets \{ d_i \mid i \in \textit{rEIds} \}$ \Comment{Retrieve data entries based on \textit{rEIds}}
    \State \Return $D$ \Comment{Return corresponding data entries}
\EndFunction

\Function{FindEntriesByBHT}{nodeId, sK, eK, rEIds, count}
    \State node $\gets$ BHashTreeNodes.nodeId
    \If{node.isLeaf}
        \If{node.isHashNode}
            \ForAll{key $\in$ $\langle$\textit{sK}, \textit{eK}$\rangle$}
                \ForAll{entryId $\in$ node.hashEty.key}
                    \State rEIds $\gets$ rEIds $\cup$ \{entryId\}, \text{count} $\gets$ \text{UpdateCount}(\text{count})
                \EndFor
            \EndFor
        \Else
            \ForAll{$\langle$key, entryId$\rangle$ $\in$ $\langle$node.keys, node.dEIds$\rangle$}
                \If{key $\geq$ sK \textbf{and} key $\leq$ eK}
                    \State rEIds $\gets$ rEIds $\cup$ \{entryId\}, \text{count} $\gets$ \text{UpdateCount}(\text{count})
                \EndIf
            \EndFor
        \EndIf
    \Else
        \ForAll{childId $\in$ node.children}
            \If{|$\mathcal{R}$(\text{Child}) $\cap$ [\textit{sK}, \textit{eK}]| > 0}
                \State count $\gets$ \Call{FindEntriesByBHT}{childId, sK, eK, rEIds, count}
            \EndIf
        \EndFor
    \EndIf
    \State \Return count
\EndFunction
\end{algorithmic}
\end{algorithm}

\subsection{Verifiable High-Performance Trie}
\label{sec:trie}
To support fuzzy queries, we present a verifiable high-performance trie based on previous work~\cite{fredkin1960trie}. 
Implementing a fast and gas-optimized trie in Solidity presents several challenges: (1) Minimizing gas consumption due to storage operations; (2) Handling timestamp strings efficiently despite lack of native string manipulation or character indexing; (3) Optimizing storage to reduce costs associated with mappings and dynamic arrays; (4) Ensuring scalability and performance given maximum call stack depth of 1024. 
To overcome these challenges, we reduce gas costs by using gas-efficient data structures and index-based storage. 
This involves assigning valid characters to fixed integer indices and storing nodes in mappings instead of arrays.
By encoding each character as an integer, we enable efficient traversal and storage without the overhead of string operations.
Furthermore, we optimize storage by replacing fixed-size arrays with mappings for child node indices in each trie node.
We enhance scalability by redesigning recursive functions into iterative ones, thus avoiding stack depth limitations and handling larger datasets efficiently within the EVM constraints.

To make the trie verifiable, we integrate hash digests into each node, where each digest is computed using a one-way hash function over the concatenation of the node's key and the digests of its child nodes.
This structure allows each node to verify the integrity of its data and its associated branches. 
The root node contains a global digest that serves as an anchor, updated with each structural modification (e.g., insertions or deletions) to maintain consistent integrity checks. 
By storing this root digest on-chain, clients can verify the correctness of any queried data without accessing the full trie, relying instead on a VO containing the path of digests from the root to the queried node along with sibling digests.

\subsection{Correctness and Completeness Analysis}
\label{sec:CorrectnessAnalysis}
\textbf{Theorem.} \emph{No false result is produced if the blockchain, off-chain storage, and verifiable indexing structures are functioning correctly.}

\noindent \textit{Proof.} 
We prove this theorem by contradiction. 
Suppose a false result is produced. 
This could only occur if (i) the verifiable indexing structures failed, and (ii) the Bloom Filter-based cache returned a false result. 
However, (i) is impossible because the BHashTree and trie structures use cryptographic digests to ensure correctness.
If any modification or tampering occurs, it would result in a mismatch of digests during verification, thereby preventing false results from being returned. 
(ii) is impossible because the Bloom Filter cache, while possibly producing false positives, cannot return false negatives.
This means that if a result is not cached, it will be explicitly retrieved from the verifiable indexing structures, where the cryptographic verification ensures correctness.

\noindent \textbf{Theorem.} \emph{No valid result is missing if the blockchain, off-chain storage, and verifiable indexing structures are functioning correctly.}

\noindent \textit{Proof.} 
We prove this theorem by contradiction. Suppose a valid result is missing. 
This could only occur if (i) the verifiable indexing structures failed, (ii) the bloom-filter-based cache excluded it incorrectly, or (iii) the verification process rejected it. 
However, (i) is impossible because the BHashTree and trie structures maintain integrity through cryptographic digests, ensuring accurate indexing. 
(ii) is impossible because the Bloom Filter cache only risks false positives, never false negatives.
(iii) is impossible because any tampering would be detected via mismatched digests during verification. 
Therefore, no valid result will be missing.

\subsection{Security and Complexity Analysis}
\label{sec:SecurityAnalysis}

\textbf{Theorem.} 
\emph{Querying operations in \Chain are secure if the blockchain, off-chain storage, and verifiable indexing structures are secure.}

\noindent \textit{Proof.} 
We prove this theorem also by contradiction. 
Suppose the querying operations in \Chain are not secure. 
This would imply (i) the middleware introduces a vulnerability, (ii) the BHashTree or trie compromises security, or (iii) the verification process fails. 
However, (i) is impossible as \Chain adheres to the blockchain's security protocols. 
(ii) is impossible because the verifiable structures use cryptographic digests to ensure integrity. 
(iii) is impossible because mismatched digests during verification would detect any tampering. 
Thus, \Chain is secure if the underlying systems are secure.

\noindent \textbf{Complexity Analysis.} 
The BHashTree structure demonstrates an insertion time complexity of \( O(\log N) \) before conversion (when \( N < T \)) and \( O(1) \) after conversion (when \( N \geq T \)). 
The search time complexity is \( O(\log N + \frac{R}{B} ) \) before conversion and \( O(R) \) after conversion, where \( R \) is the number of result entries and \( B \) is the branching factor. The space complexity remains \( O(N) \). 
For the high-performance trie structure, the construction time complexity is \( O(n \, \overline{l}) \), where \( n \) is the number of strings and \( \overline{l} \) is the average string length. 
Its space complexity is \( O(n \, \overline{l} \, \sigma) \) in the worst case (where \( \sigma \) is the alphabet size) but is typically \( O(n \, \overline{l}) \) in practical scenarios.
Query operations have a time complexity of \( O(l) \), with \( l \) being the length of the query string, and a space complexity of \( O(1) \).

\section{Experimental Design}

We present the experimental design, including studied datasets, baselines, evaluation metrics, and experiment settings.

\subsection{Datasets}
\label{sec:dataset}

We construct two datasets: one Ethereum dataset (abbreviated as ETH)~\cite{awesomeBigquery2022} and one Bitcoin dataset (abbreviated as BTC)~\cite{vanclassifying}. 
After necessary pre-processing and sampling, our dataset contains 16,384 blocks, with transactions reorganized into the format \( O_i = \langle A_i, a_i, t_i \rangle \), where \( A_i \) represents the amount, \( a_i \) represents the set of addresses involved, and \( t_i \) represents the timestamp.
To create a multimodal dataset, we incorporate an \textit{imagecid} or \textit{videocid} value into the formatted transactions, establishing a link between on-chain transactions and off-chain data. 
The final structure is a quintuple \( O'_i = \langle A_i, a_i, t_i, \textit{imagecid}_i, \textit{videocid}_i \rangle \), where one of the \textit{cid}s may be null. Here, \(\textit{imagecid}_i\) and \(\textit{videocid}_i\) represent the content identifiers (CIDs) for images and videos, respectively. 
The image dataset consists of 10,000 art images in the 24x24, 8-bit style of CryptoPunk, composed of irregular pixels, while the video dataset includes 100 videos from opensea.io, a video NFT platform.


\subsection{Baselines and Evaluation Metrics}

{\bf Studied Baselines.} 
Many studies have examined verifiable query processing over blockchain databases. 
vChain+~\cite{wang2022vchain+} employs cryptographic set accumulators to enable verifiable boolean range queries.
Therefore, we use vChain+ as the baseline and replicate vChain+.
For a fair comparison, we add an off-chain query module with IPFS to make vChain+ capable of off-chain queries.
The on-chain-only and full version is denoted as vChain+$_O$ and vChain+$_F$, respectively.







\noindent
{\bf Evaluation Metrics.} 
We evaluate the performance of each workload by breaking it down into (1) {Index Construction Cost}, the CPU time cost to insert multimodal data into the blockchain and off-chain storage; (2) {Query Latency}, the time cost for querying metadata from the blockchain and multimodal data from off-chain storage; and (3) {VO Size}. Based on the Awesome BigQuery Views project\cite{awesomeBigquery2022}, we designed six types of SQL Queries to test \Chain.

\subsection{Implementation}
\label{sec:setting}

We experiment on a Ubuntu 22.04 workstation equipped with an Intel\(^\text{\textregistered}\) Core\texttrademark\ i7-12700KF Processor and 48 GB memory. 
In the experiment, we deploy a private Ethereum network with Geth and a FISCO BCOS permissioned chain, configured with nodes in a local network with a speed of 100 Mbps. 
We reimplement vChain+, the SOTA open-source system in the on-chain query domain, as our baseline (labelled as vChain+$_{O}$ and vChain+$_{F}$). \Chain is implemented with smart contracts Solidity and deployed with web3.py library in Python.
For each experiment, we randomly generate 2,046 queries based on preset templates and report average results for all metrics.

\section{Experimental Results}

To investigate the performance of \Chain, our experiments focus on the following three research questions:

\begin{itemize}[leftmargin=*]
\item \textbf{RQ-1 Comparable Study on Time Range Query.} {\em How does the performance of ~\Chain compare with the baselines on Time Range Query?}

\item \textbf{RQ-2 Comparable Study on Fuzzy Query.} {\em How does the performance of ~\Chain compare with the baselines on Fuzzy Query?}

\item \textbf{RQ-3 Gas Efficiency and Scalability Analysis.} {\em How does \Chain perform in terms of gas consumption and scalability compared to the baselines?}
\end{itemize}

\subsection{RQ-1 Comparable Study on Range Query}

\noindent
\textbf{Objective.}
Benefiting from the powerful query capability of Btree structures, many blockchain range query approaches have been proposed~\cite{xu2019vchain,wang2022vchain+}.
vChain and vChain+ are verifiable blockchain query systems that support range query, demonstrating excellent performance across various range query tasks.
The experiments are conducted to investigate whether MulChain outperforms SOTA blockchain range query approaches.

\noindent
\textbf{Experimental Design.}
We consider three baselines in our ablation study: 
vChain+$_{O}$, vChain+$_{F}$, and MulChain$_{BT}$. 
Here, vChain+$_{F}$ represents the vChain+ replication version to support on/off-chain queries, referred to as vChain+$_{FB}$ for Bitcoin and vChain+$_{FE}$ for Ethereum.
vChain+$_{O}$ is vChain+$_{F}$ without the off-chain query module, labeled as vChain+$_{OB}$ and vChain+$_{OE}$.
MulChain$_{BT}$ is MulChain with its underlying data structure replaced with B\(+\)Tree for time range queries, represented as MulChain$_{BTB}$ and MulChain$_{BTE}$.
MulChain$_{BH}$ is MulChain with its underlying data structure replaced with our gas-efficient BHashTree for time range queries, shown as MulChain$_{BHB}$ and MulChain$_{BHE}$.

\begin{figure}[htbp]
    \centering
    \includegraphics[width=\linewidth]{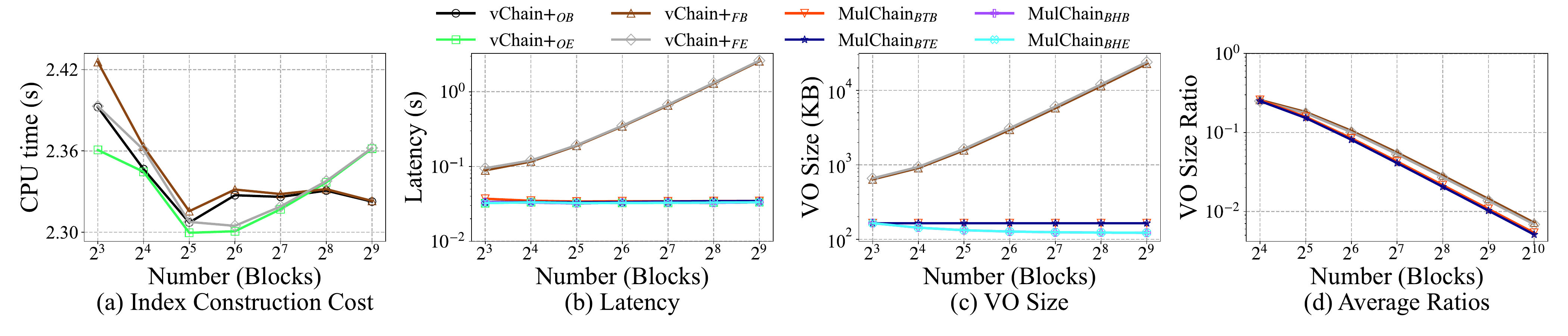}
    \caption{Time Range Query Performance}
    \label{fig:Time Range Query Performance}
\end{figure}

\noindent
\textbf{Results.}
MulChain$_{BH}$ is 78.13X faster than vChain+$_F$. 
Notably, our BHashTree takes up 196.08X less VO size than vChain+. 
That is because the accumulator employed in vChain+ takes a lot of storage compared with the mapping structure used in MulChain. 
From Fig.~\ref{fig:Time Range Query Performance}(a), the insertion overhead for time range queries generally follows a similar trend to simple queries. 
Also, the time cost of off-chain queries decreases as the block number increases thanks to the bloom filter cache~\cite{ScalableBloomFilters2007}.
Fig.~\ref{fig:Time Range Query Performance}(c) shows a decreasing VO size for BHashTree methods as the block number increases and the VO size of B\(+\)Tree and BHashTree are identical for the first 8 blocks because BHashTree doesn't convert B\(+\)Tree structure to hash nodes before reaching insertion threshold (which is set to 10).
This result aligns with expectations, as BHashTree gradually converts B\(+\)Tree nodes (with larger VO size) to hash nodes (with smaller VO size).
Finally, Fig.~\ref{fig:Time Range Query Performance}(d) shows that as the block number increases, our tree structures consume progressively less VO size compared to vChain+.

\intuition{
{\bf Answer to RQ-1}:
MulChain$_{BH}$, utilizing the gas-efficient BHashTree, is 78.13X faster than vChain+$_{F}$, with a significant reduction in VO size—196.08X smaller than vChain+. The BHashTree's structure is more storage-efficient compared to vChain+'s accumulator, and its VO size decreases as the block number increases due to the conversion of B+Tree nodes to hash nodes. These results highlight the superior performance and storage efficiency of MulChain$_{BH}$ for time range queries.
}

\subsection{RQ-2 Comparable Study on Fuzzy Query}

\noindent
\textbf{Objective.}
Benefiting from the powerful query capability of trie structures, many fuzzy query approaches have been proposed~\cite{tahani1977conceptual,fagin1998fuzzy}.
However, no such approaches support blockchain-level fuzzy queries. 
To mitigate that, we propose a verifiable trie structure in Section~\ref{sec:trie}.
The experiments are conducted to investigate whether \Chain achieves a considerable fuzzy query performance on blockchain databases.

\noindent
\textbf{Experimental Design.}
We consider two baselines in our ablation study: 
vChain+$_{O}$ and vChain+$_{F}$.
Here, vChain+$_{F}$ represents the vChain+ replication version to support on/off-chain queries, referred to as vChain+$_{FB}$ for Bitcoin and vChain+$_{FE}$ for Ethereum.
vChain+$_{O}$ is vChain+$_{F}$ without the off-chain query module, labeled as vChain+$_{OB}$ and vChain+$_{OE}$.
MulChain$_{T}$ is MulChain with its underlying data structure replaced with our verifiable trie for fuzzy queries, identified as MulChain$_{TB}$ and MulChain$_{TE}$.





\begin{figure}[htbp]
    \centering
    \includegraphics[width=\linewidth]{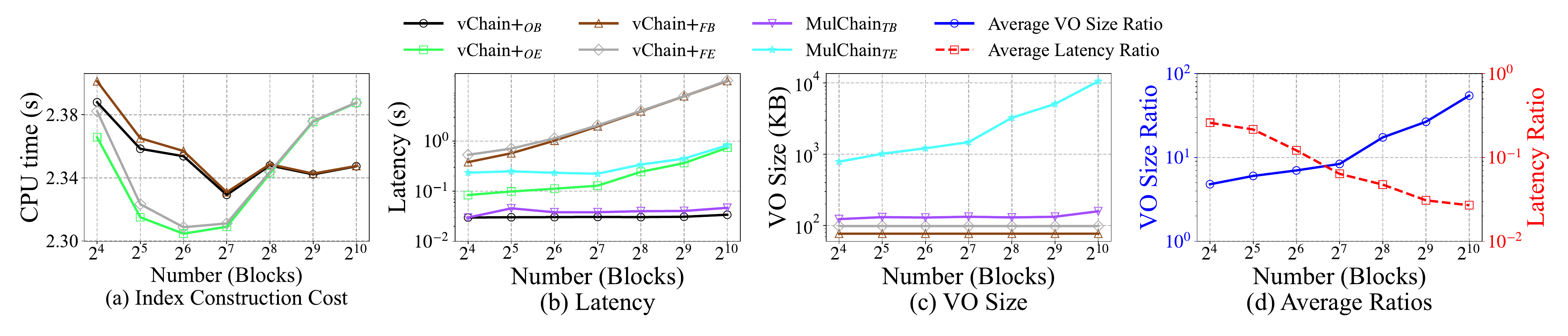}
    \caption{Fuzzy Query Performance}
    \label{fig:Fuzzy Query Performance}
\end{figure}

\noindent
\textbf{Results.}
MulChain$_T$ is 30X faster than vChain+$_F$.
As shown in Fig.~\ref{fig:Fuzzy Query Performance}(a), the insertion overhead aligns with that of the other two query types, which is expected since datasets are inserted uniformly across all three queries. Fig.~\ref{fig:Fuzzy Query Performance}(b) indicates that the accumulator-based MulChain$_V$ exhibits an exponential growth trend in the number of blocks. At the same time, our trie demonstrates \( O(l) \) time complexity. Notably, fuzzy query support was absent in vChain+ before our reimplementation. The query latency for the ETH dataset is significantly higher than that for the BIT dataset. This is due to the denser timestamp distribution in the ETH dataset. We intentionally configured this distribution to test fuzzy queries under varying workload intensities. The average latency of MulChain$_V$ is 30X higher than that of MulChain$_T$, highlighting the superiority of the trie over the accumulator-based approach. In Fig.~\ref{fig:Fuzzy Query Performance}(c), it is evident that our trie method consumes 10.7X more VO size than the accumulator-based method in the worst case. However, this outcome is anticipated as we achieve an acceptable trade-off between VO size and latency, as shown in Fig.~\ref{fig:Fuzzy Query Performance}(d).

\intuition{
{\bf Answer to RQ-2}: 
MulChain$_{T}$, using the verifiable trie, is 30X faster than vChain+$_{F}$ for fuzzy queries, demonstrating \( O(l) \) time complexity, compared to the exponential growth of the accumulator in vChain+$_{F}$. 
Although it incurs 10.7X more VO size in the worst case, this trade-off is acceptable due to significantly reduced query latency. 
The trie outperforms other approaches, especially under varying workload intensities, making it a superior choice for blockchain fuzzy queries.
}

\subsection{RQ-3 Gas Efficiency and Scalability Analysis}
\label{sec:rq3}

\noindent
\textbf{Objective.}
In \Chain, we take \textit{Gas Fee} and \textit{Scalability} into consideration.
In blockchain query databases, the gas fee and scaleability are essential since the former will largely impact the practicality of the designed methods (i.e., a higher gas fee means more money spent when running queries) and the latter will impact compatibility (i.e., poor compatibility will lead to massive code modification when adjusting to another blockchain system).

\noindent
\textbf{Experimental Design.}
First, we investigate the impact of different data structures on gas fee and design four variants of \Chain. 
vChain+$_{F}$ represents vChain+ replicated and enhanced to support multimodal queries.
vChain+$_{O}$ is vChain+$_{F}$ without the off-chain query module.
MulChain$_{BT}$ is MulChain with its underlying data structure replaced with B\(+\)Tree for time range queries.
MulChain$_{BH}$ is MulChain with its underlying data structure replaced with our gas-efficient BHashTree for time range queries.
MulChain$_{T}$ is MulChain with its underlying data structure replaced with our verifiable trie for fuzzy queries.
This approach allows us to examine the individual effects of each component.

\noindent
\textbf{Results.} We discuss the results from the aspects of gas consumption and scalability, respectively.

\begin{figure}[htbp]
    \centering
    \includegraphics[width=.7\linewidth]{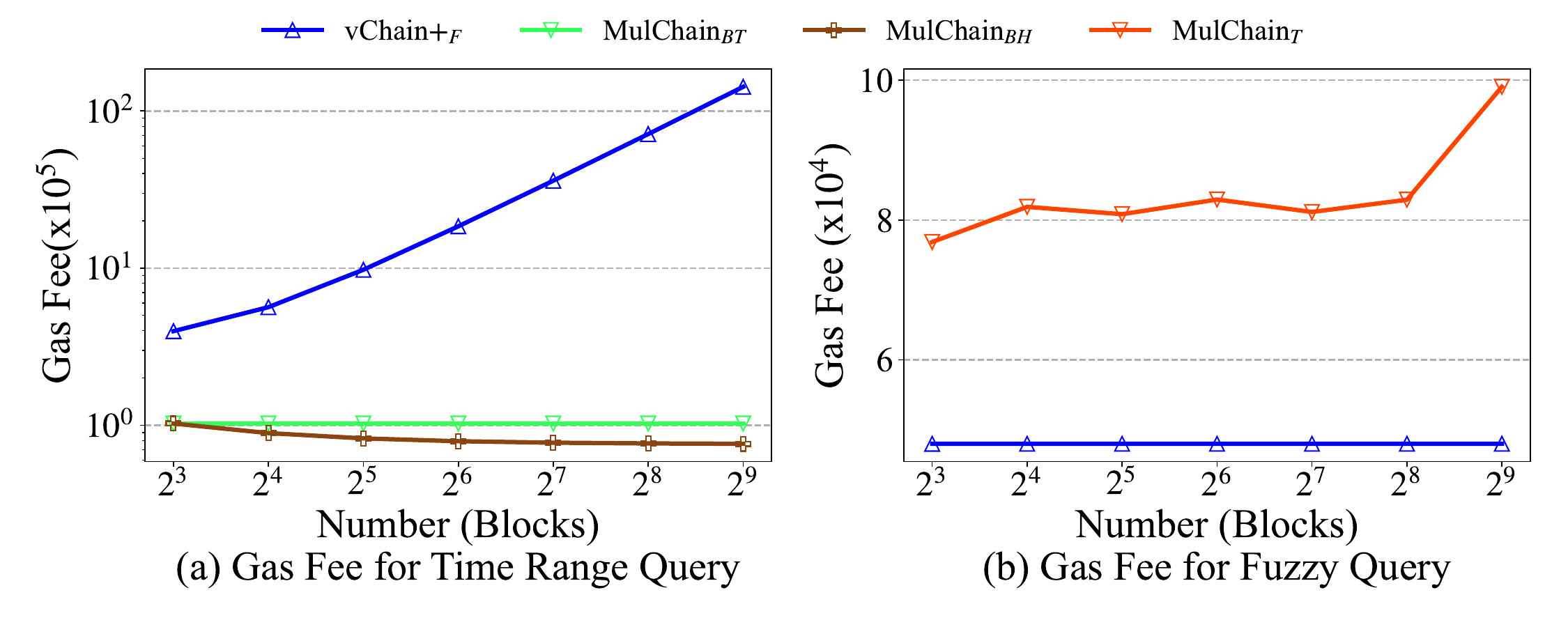}
    \caption{Gas Consumption}
    \label{fig:Gas Consumption}
\end{figure}

\noindent
\textbf{\underline{Gas Consumption Analysis.}}
The average gas fees for BHashTree are much lower than those of vChain+$_{F}$, slightly lower than B\(+\)Tree-based methods, as illustrated in Fig.~\ref{fig:Gas Consumption}(a).
Fig.~\ref{fig:Gas Consumption}(b) presents the average gas fees of the trie in comparison to the accumulator from vChain+. 
Notably, the gas consumption of MulChain$_{T}$ exceeds that of vChain+$_{F}$ due to our strategic trade-off of space for time. 
We deem this trade-off acceptable, as the reduction in query latency is particularly valuable in the context of fuzzy queries on blockchains.

\begin{figure}[htbp]
    \centering
    \includegraphics[width=.7\linewidth]{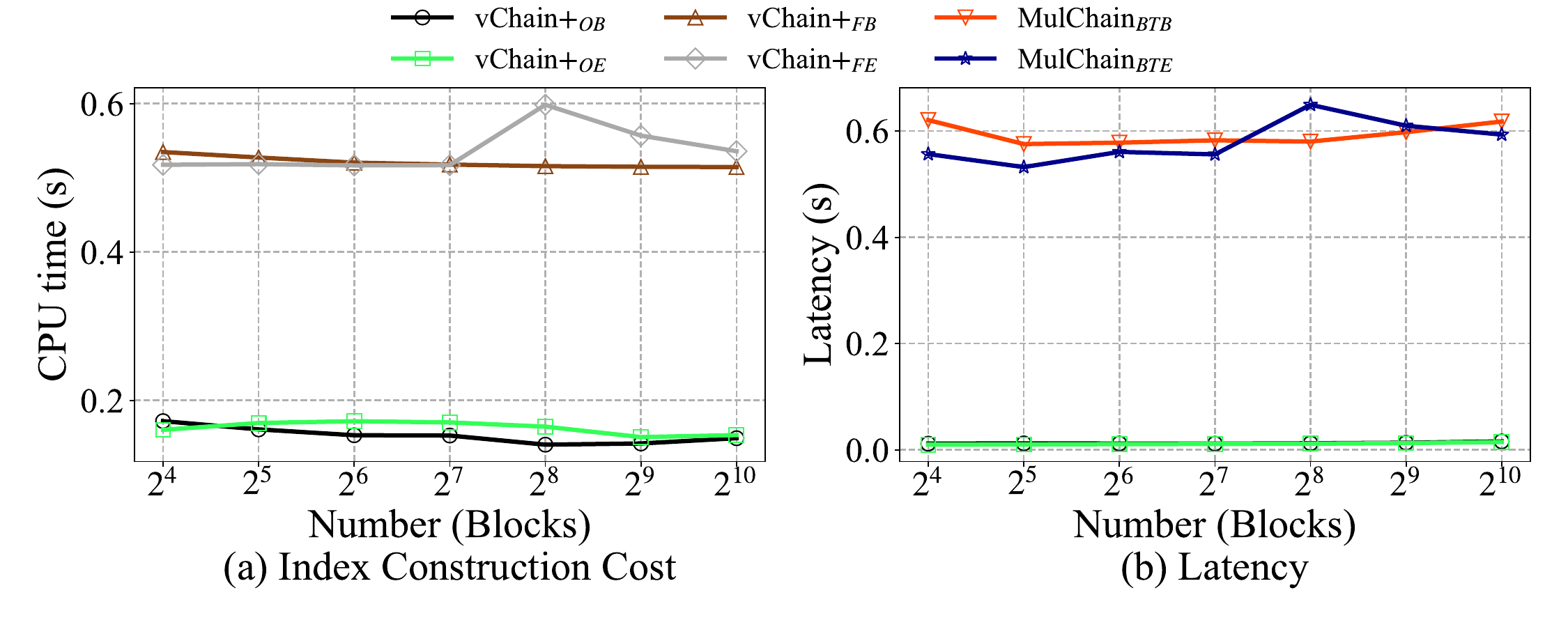}
    \caption{Query Performance on FISCO BCOS}
    \label{fig:FISCO BCOS}
\end{figure}

\noindent
\textbf{\underline{Scalability Analysis.}}
\Chain supports all six SQL primitives (i.e., insert, delete, update, simple, time range, fuzzy queries) on Ethereum and FISCO BCOS. 
In contrast, the CRUD Service of FISCO BCOS does not support time range and fuzzy queries. 
We test \Chain using time range queries on BTC and ETH datasets. 
From Fig.~\ref{fig:FISCO BCOS}(a), we can see that \Chain undergoes a decline of up to 3.78\% when the number of blocks grows.
In Fig.~\ref{fig:FISCO BCOS}(b), we observe that MulChain$_{BT}$ is faster on the BTC dataset than on ETH for block counts below 128 and above 1024. 
This performance difference is due to the varying timestamp densities of the two datasets and the initialization cost of the B\(+\)Tree.

\intuition{
{\bf Answer to RQ-3}: 
(1) The five data structures (i.e., accumulator of vChain+$_{F}$, vChain+$_{O}$, MulChain$_{BT}$, MulChain$_{BH}$ and MulChain$_T$) contribute substantially to \Chain, and combining them achieves the best performance of blockchain query on different scenarios.
(2) The gas fee of MulChain$_{T}$ exceeds that of vChain+$_{F}$ due to our strategic trade-off of space for time.
(3) \Chain supports blockchains based on Ethereum virtual machine and Hyperledger Fabric.
}



\section{Related Work}

Table~\ref{tab:Comparison} compares various query processing systems. These systems are classified into two categories: {outsourced databases} and {blockchain databases}.

\subsection{Outsourced Databases}
Numerous systems have been developed to tackle the challenges of data integrity, security, and query verifiability in cloud-based outsourced databases. IntegriDB~\cite{zhang2015integridb} addresses data integrity by allowing clients to verify the correctness of query results from outsourced databases, but it supports only a limited subset of SQL queries. FalconDB~\cite{peng2020falcondb} enhances verifiable range queries over outsourced data, providing improved query capabilities; however, it faces scalability issues when dealing with large datasets, which hampers its performance in big data applications. vSQL~\cite{zhang2017vsql} proposes a protocol for verifiable SQL queries on dynamic outsourced databases, offering strong security guarantees, yet it incurs substantial computational overhead, making it less suitable for real-time query processing. SQL Ledger~\cite{antonopoulos2021sql} leverages blockchain to ensure the integrity and immutability of database transactions but lacks support for rich query functionalities and efficient data retrieval mechanisms, limiting its applicability in scenarios requiring complex data analysis. LedgerDB~\cite{yang2020ledgerdb} integrates blockchain with database systems to create a tamper-proof ledger that supports query operations. GlassDB~\cite{yue2022glassdb} introduces a verifiable data structure designed for outsourced databases to enhance query verifiability and integrity. These solutions can be used for querying blockchain data in theory. However, they often struggle to support multiple query types and do not provide the efficient performance required by modern, data-intensive applications.

\subsection{Blockchain Databases}
Many studies have examined verifiable query processing over blockchain databases. vChain~\cite{xu2019vchain} and vChain+~\cite{wang2022vchain+} employ cryptographic set accumulators to enable verifiable boolean range queries. To address the issue of public key management in vChain, vChain+ introduces a sliding window accumulator (SWA) to reduce public key storage overhead. However, their design requires a large amount of storage space for verification objects (VO) and incurs significant computational overhead. Shao et al.~\cite{shao2020authenticated} proposed an authenticated range query scheme using Trusted Execution Environments (TEE). However, due to limited secure memory space, existing TEEs struggle to handle large-scale applications. GEM$^2$-tree~\cite{zhang2021authenticated,GEM^2-Tree} offers a gas-efficient two-level index structure that supports authenticated queries, aiming to lower on-chain storage costs in Ethereum. However, maintaining the tree-based authenticated data structure (ADS) imposes a heavy burden on light nodes during verification procedures. LVQ~\cite{dai2020lvq} targets Bitcoin transaction analysis and uses a Bloom filter-integrated authenticated index~\cite{ScalableBloomFilters2007} to verify transaction membership. Although it provides lightweight verifiability, it does not support time-range queries. These approaches often limit query types due to their specialized index designs and typically consider a single blockchain data source. Moreover, many of these solutions require custom blockchain structures, making them incompatible with existing blockchain networks. The Graph (TG)~\cite{thegraph2022} is a decentralized protocol designed for indexing blockchain data. It employs indexer nodes to aggregate information from multiple blockchains, thereby offering flexible and efficient query services. To ensure the reliability of query results, TG implements a dispute resolution mechanism. However, this mechanism does not guarantee query integrity and may cause significant delays during verification. V\textsuperscript{2}FS~\cite{V2FS2024} proposes a virtual filesystem that facilitates verifiable query processing over multi-chain data by shifting the focus from verifying computation to verifying data, enabling efficient integrity assurance using an off-the-shelf database engine. However, most of these methods require massive modifications to the underlying blockchain codes. This limits the generalizability and versatility of these methods. \Chain differs from these and does not need any change to the blockchain, thus can be integrated with various blockchains and off-chain storage solutions. 

In summary, existing systems cannot support all desired features, including advanced multimodal queries, blockchain compatibility and off-chain compatibility. In contrast, \Chain is the first to cover all of these features simultaneously.

\section{Threats to Validity}

\noindent
\textbf{Internal Validity.}
One potential threat to the internal validity of this study is the experimental setup, including the specific configurations of the blockchain platforms and off-chain storage systems. The performance of MulChain may vary across different network setups and configurations, particularly in terms of query latency and gas consumption. The experiments are based on a specific testbed (Ethereum and FISCO BCOS), and the results may not fully generalize to other blockchain systems or hybrid-storage solutions. Furthermore, the middleware relies on the assumption that the underlying blockchain systems and storage solutions remain stable and function as expected, which may not always hold true in more dynamic or resource-constrained environments.
Another threat is the design of the proposed BHashTree and trie structures. While these indexing structures are presented as being gas-efficient and verifiable, their performance in extremely high-load scenarios, with large-scale multimodal data sets, remains an open question. The dynamic adaptation mechanism for BHashTree, which transitions between B+Tree and hash nodes, assumes predictable query and insertion patterns. If these patterns do not align with the assumed behavior, the performance benefits might be diminished. Additionally, the verifiability and security of the trie structure depend on the correctness of the cryptographic digests and the integrity of the underlying data.

\noindent
\textbf{External Validity.}
The external validity of the study is potentially limited by the focus on specific datasets (Ethereum and Bitcoin) and particular blockchain platforms (Ethereum and FISCO BCOS). While the proposed MulChain middleware is designed to be compatible with a variety of blockchain systems, the study does not explore the full range of blockchain environments or data types that might affect the performance and scalability of the system. For example, the characteristics of the Ethereum and FISCO BCOS networks, such as transaction volume, block size, and consensus mechanisms, may influence the applicability of MulChain to other blockchain networks with different characteristics, such as Hyperledger Fabric or newer blockchain systems.
Moreover, the study is based on the assumption that the integration of off-chain storage (such as IPFS) is suitable for all use cases involving multimodal data. However, the suitability of IPFS for specific types of data or scenarios—such as real-time data retrieval or high-security applications—could be questioned. In real-world decentralized applications (DApps), the interaction between on-chain and off-chain components could be more complex and subject to issues such as data integrity, privacy concerns, or network latency, which might impact the generalizability of the findings.

\section{Conclusion}
\label{sec:conclusion}

\noindent In this paper, we propose \Chain, the first work to enable advanced cross-modal queries on hybrid-storage blockchains. 
To achieve that, two indexing structures integrating a verifiable gas-efficient BHashTree and a high-performance trie are proposed. 
BHashTree supports transitions between a B\(+\)Tree and a hashtable based on workload type. 
Our system seamlessly integrates with existing blockchains and off-chain storage solutions. 
Experimental results show that \Chain achieves up to 78.12X speedup in query performance and reduces the VO size by 99.49\% in comparison with the state-of-the-art vChain+.
Several interesting problems merit further investigation, such as supporting more complex cross-modal queries and optimizing the performance of our methods.

\section*{Acknowledgements}
This work was supported in part by National Key Research and Development Program of China under Grant 2024YFB2705300, in part by the National Natural Science Foundation of China (NSFC) under Grant 62402313, in part by the Shanghai Science and Technology Innovation Action Plan under Grant 23511100400, in part by the Open Research Fund of The State Key Laboratory of Blockchain and Data Security, Zhejiang University.

\balance
\bibliographystyle{ACM-Reference-Format}
\bibliography{main}

\end{document}